\documentclass[sigconf,natbib=false]{acmart}

\RequirePackage[
  datamodel=acmdatamodel,
  style=acmnumeric,
  ]{biblatex}

\copyrightyear{2026}
\acmYear{2026}
\setcopyright{cc}
\setcctype{by}
\acmConference[DAC '26]{63rd ACM/IEEE Design Automation Conference}{July 26--29, 2026}{Long Beach, CA, USA}
\acmBooktitle{63rd ACM/IEEE Design Automation Conference (DAC '26), July 26--29, 2026, Long Beach, CA, USA}
\acmDOI{10.1145/3770743.3804202}
\acmISBN{979-8-4007-2254-7/2026/07}

\usepackage{url}
\usepackage{xspace}
\usepackage{multirow}
\usepackage{pifont}
\usepackage{xcolor}

\usepackage{tikz}

\newcommand*{\circled}[1]{\tikz[baseline=(char.base)]{
    \node[shape=circle, fill=black, inner sep=2pt] (char) {\textcolor{white}{#1}};}}

\newcommand{\hackthesilicon}{HackTheSilicon\xspace} 

\newcommand{\htsparticnt}{1,600}

\newcommand{\htscnt}{15}

\newcommand{\ournameNoSpace}{AttackonCTF} 
\newcommand{\ourname}{\ournameNoSpace\xspace}

\begin{document}

\title[AttackonCTF]{AttackonCTF: Defending Hardware Security Competition Benchmarks in the Age of LLMs}

\author{Mohamadreza Rostami$^\dagger$, Nikhilesh Singh$^\dagger$, Stephen Muttathil$^\S$, Lichao Wu$^\ddagger$, Chen Chen$^\S$, Huimin~Li$^\diamond$, Jeyavijayan Rajendran$^\S$ and Ahmad-Reza Sadeghi$^\dagger$}

\affiliation{$^\dagger$Technical University of Darmstadt\country{Germany}\>\> $^\S$Texas A\&M University\country{USA}\>\> \\ $^\ddagger$University of Bristol\country{United Kingdom}\>\> $^\diamond$Shenzhen University of Advanced Technology\country{China} \\ \normalsize \texttt{\{mohamadreza.rostami,nikhilesh.singh,ahmad.sadeghi\}@trust.tu-darmstadt.de \{chenc,stephen7929,jv.rajendran\}@tamu.edu\>\> lichao.wu@bristol.ac.uk\>\> lihuimin@suat-sz.edu.cn}}

\renewcommand{\shortauthors}{Rostami et al.}

\begin{abstract}

Hardware security competitions such as HackTheSilicon serve as benchmarking platforms for evaluating vulnerability detection methods and for training humans and AI. However, our study reveals that LLMs threaten their validity. Instead of genuine security reasoning, detectors exploit a diff-style syntactic comparison, achieving an 83\% detection rate, undermining fair evaluation. To mitigate this, we propose the first LLM-oriented, semantics-preserving obfuscation framework for these benchmarks. Unlike IP-protection approaches, it applies human-readable transformations and controlled diff-noise while preserving functionality. On HackTheSilicon, the framework reduces LLM-based detection accuracy by 50\% with only 10\% obfuscation and by 78.6\% under complete obfuscation, restoring benchmark reliability.

\end{abstract}



\maketitle

\section{Introduction}
\label{sec:introduction}
The exponential growth in hardware complexity has fundamentally transformed hardware security~\cite{dessouky2019hardfails,rostami2024fuzzerfly, akter2023survey, koylu2023survey}. However,
the predominantly black-box nature of hardware creates a critical gap between industry needs and academic research capabilities~\cite{hackthesilicon, dessouky2019hardfails}. Indeed, commercial hardware designs are typically proprietary, limiting researchers' access to realistic security challenges, thus creating a development and validation gap for vulnerability detection techniques~\cite{marks2025ResearchGaps,hackthesilicon, specure, dessouky2019hardfails}.
To address this challenge, we organize the \hackthesilicon event in partnership with Intel and Synopsys~\cite{hackthesilicon}. Initiated in 2018, we have hosted \htscnt{} competitions with over \htsparticnt{} participants, using open-source SoC designs embedded with industry-informed vulnerabilities~\cite{sadeghi2021organizing, dessouky2019hardfails, chen2022trusting, 10.1145/3316781, hackthesilicon}. \hackthesilicon serves two primary purposes: (1) providing industry-validated open-source vulnerability benchmarks for evaluating existing or newly developed vulnerability detection tools, and (2) training hardware security researchers through realistic challenges. Furthermore, these benchmarks serve broader purposes: evaluating LLM-based vulnerability detection approaches~\cite{collini2025marvel}, training AI models for hardware security analysis~\cite{collini2025marvel}, and serving as demonstrative examples in the Common Weakness Enumeration (CWE)~\cite{cwehwtopten} database.

\noindent\textbf{``Shortcut'' with LLM.} 
\hackthesilicon is built on open-source SoC designs with injected vulnerabilities~\cite{hackthesilicon,dessouky2019hardfails, Kanuparthi2024HackAtDAC}. Due to the availability of the register-transfer level (RTL) code, participants or detection tools could, in principle, simply diff the original and modified SoCs to locate inserted bugs without performing any real security analysis. To prevent this, the bug submission process has always required a comprehensive technical report for each identified vulnerability, including its security impact, CWE/CVSS classifications, and proof-of-concept exploits. This requirement ensures that participants cannot rely solely on diffing and must demonstrate genuine hardware security understanding.

However, recent advances in LLMs~\cite{matarazzo2025survey} undermine this protection mechanism~\cite{collini2025marvel,ahmad2025lashed}. Specifically, LLM-based vulnerability detection tools can bypass the intended need for deep vulnerability analysis by combining simple diffing with LLM augmentation techniques, such as retrieval-augmented generation (RAG) that leverage prior open-source designs, history branches, and commits~\cite{opentitan,balkind2016openpiton}. Importantly, while RAG may appear to perform genuine analysis, it actually retrieves the original design context (based on similarity). This process of fetching similar code is functionally equivalent to an automated diff against the buggy code in the prompt. As a result, these methods generate high-quality, \emph{seemingly} expert vulnerability reports without performing genuine security reasoning on the vulnerable design. While they may perform well in CTF settings, we argue that they threaten both the educational integrity of such competitions and the validity of benchmarks meant to evaluate LLM-based methods that are expected to reason directly on the malicious design. Indeed, high performance may simply reflect automated diffing either explicitly or implicitly via RAG and text generation rather than real analysis.

\noindent\textbf{Our Objective \& Solution.} Given the effectiveness of LLM-based detection methods, a natural question arises: \emph{How can we prevent existing LLM-based ``shortcuts'' while maintaining the educational and research integrity of the CTF?} While potential solutions exist, such as patch randomization, structural refactoring, and multi-branch reshuffling, an ideal solution must be lightweight enough for practical use, aligned with the objectives of \hackthesilicon, and, crucially, must preserve the ability of participants and researchers to learn from and reason about the malicious design. This leads to a natural alternative: \emph{hardware obfuscation}. Traditional obfuscation tools~\cite{aldecHdlObfuscation, semanticdesigns_verilog_obfuscator, verible} were developed for IP protection and rely on aggressive techniques, e.g., identifier encryption, hashing, and layout anonymization, that render RTL extremely challenging to understand. While ideal for preventing reverse engineering, as shown in Table~\ref{tab:comparison}, this level of obfuscation directly conflicts with the goals of a security competition or open benchmark: it prevents human learning, hinders the development of next-generation LLM-based detectors, and eliminates the usefulness of examples for CWE-style vulnerability documentation. 

\newcommand{\cmark}{\textcolor{green}{\ding{51}}}%
\newcommand{\xmark}{\textcolor{red}{\ding{55}}}%

\begin{table}[t]
\centering
\footnotesize
\caption{Traditional Obfuscation vs. \ourname}
\label{tab:comparison}
\begin{tabular}{lcc}
\hline
\textbf{Characteristic} & \textbf{Traditional} & \textbf{\ournameNoSpace} \\
\hline
Primary Goal & IP Protection & Benchmark Integrity \\
\hline
\multirow{2}{*}{Target Adversary} & Human Reverse & LLM-based \\
 & Engineers & Detectors \\
\hline
Human Readability & \xmark & \cmark \\
\hline
Educational Value & \xmark & \cmark \\
\hline
AI Training Utility & \xmark & \cmark \\
\hline
CWE Usability & \xmark & \cmark \\
\hline
\end{tabular}
\end{table}

In this work, we introduce the first LLM-oriented obfuscation framework tailored specifically for hardware security competitions and benchmarks. Our approach preserves readability while strategically disrupting the syntactic patterns and structural cues exploited by LLMs. The framework applies semantic-preserving transformations to break direct correlation with the original code and injects benign modifications to create controlled diff noise that masks the true security-relevant changes. Unlike traditional obfuscation, the resulting designs remain interpretable for human analysts and suitable for developing and evaluating (LLM-based) reasoning-based detection tools. Overall, our main contributions are:
\begin{itemize}
\item We provide the first empirical demonstration that LLM-based vulnerability detection benefits heavily from diffing and syntactic comparison, achieving 75–83\% detection rates without genuine hardware security reasoning.
\item We develop the first readability-preserving, LLM-specific obfuscation framework with diff noise, context dilution, and semantic misdirection, effectively defending diff-based LLM exploitations.
\item On \hackthesilicon benchmarks, our method reduces the detection accuracy of the existing approach by 50\% at 10\% obfuscation and by 78.6\% at full obfuscation, without compromising human readability or functional correctness.
\item Our format-agnostic framework incorporates automated verification pipelines to ensure that all obfuscated designs remain semantically equivalent and fully functional.
\end{itemize}

\section{Background}
\label{sec:background}

Unlike the software community, where numerous open-source applications exist, including widely-used and commercialized systems such as the Linux kernel, which serve as benchmarks for security analysis and evaluation of vulnerability detection tools, the hardware community faces significant limitations. Due to the proprietary nature of hardware intellectual property (IP), open-source hardware designs are scarce, with only a limited number of projects available~\cite{balkind2016openpiton, opentitan}. This scarcity creates a substantial gap between the quality of security analysis requirements and standards maintained in industry versus those accessible to academia.

To address this gap, initiatives combining academia and industry expertise have emerged. \hackthesilicon (also known as Hack@Event), founded in 2017, has become the world's largest hardware security competition~\cite{hackdac_origin}. 
Beyond their educational value in training the next generation of hardware security researchers, these competitions serve as critical benchmarking platforms for the hardware security research community, providing standardized vulnerability datasets that enable researchers to evaluate and compare automated vulnerability detection tools and methodologies against industry-relevant security flaws.

These competitions follow a structured format. First, organizers select complex open-source SoC designs such as OpenPiton~\cite{balkind2016openpiton} or OpenTitan~\cite{opentitan} and inject vulnerabilities inspired by real-world vulnerabilities, guided by insights from industry partners and CWE classifications~\cite{mitre}. Vulnerabilities span various categories, including cryptographic weaknesses, access control flaws, information leakage channels, and life cycle security issues, providing academia with realistic benchmarks for evaluating vulnerability detection approaches. Next, teams receive the modified SoC design and industry-standard EDA tools from Synopsys~\cite{synopsys}, simulating real-world hardware security verification workflows. Participants are free to use any tools to detect vulnerabilities and submit comprehensive reports, including vulnerability localization, technical security analysis, CWE/CVSS classifications, proof-of-concept exploits, and methodology descriptions. Finally, a team of industry and academic experts evaluates these submissions.

\section{Design and Implementation}\label{sec:design}\label{sec:method}

To scientifically validate our hypothesis that LLMs exploit syntactic difference patterns rather than performing genuine hardware security analysis, in Section~\ref{subsec:method:bugdetection}, we design four progressively sophisticated LLM-based vulnerability detection methods, inspired by recent LLM-assisted code analysis research~\cite{guo2025svagent, collini2025marvel, saha2025sv, tarek2025bugwhisperer} and attack strategies observed from \hackthesilicon participants~\cite{hackthesilicon}. 
Guided by these insights, in Section~\ref{subsec:method:obfuscation}, we develop an LLM-driven obfuscation framework to break these detection mechanisms while preserving readability and functional correctness.

\subsection{LLM-Based Vulnerability Detection Methods}\label{subsec:method:bugdetection}

We consider four vulnerability detectors (\textbf{D1}-\textbf{D4}) with progressively richer contextual awareness to systematically characterize the LLM-assisted hardware vulnerability detection methods. These detectors represent realistic attack scenarios that competition participants or automated tools could employ, ranging from naive diff-based approaches to sophisticated systems incorporating repository history and domain knowledge. We standardize the prompting interface, decoding parameters, and output schema to isolate the effect of context available to the model. 

\begin{figure}
    \centering
    \includegraphics[width=\linewidth]{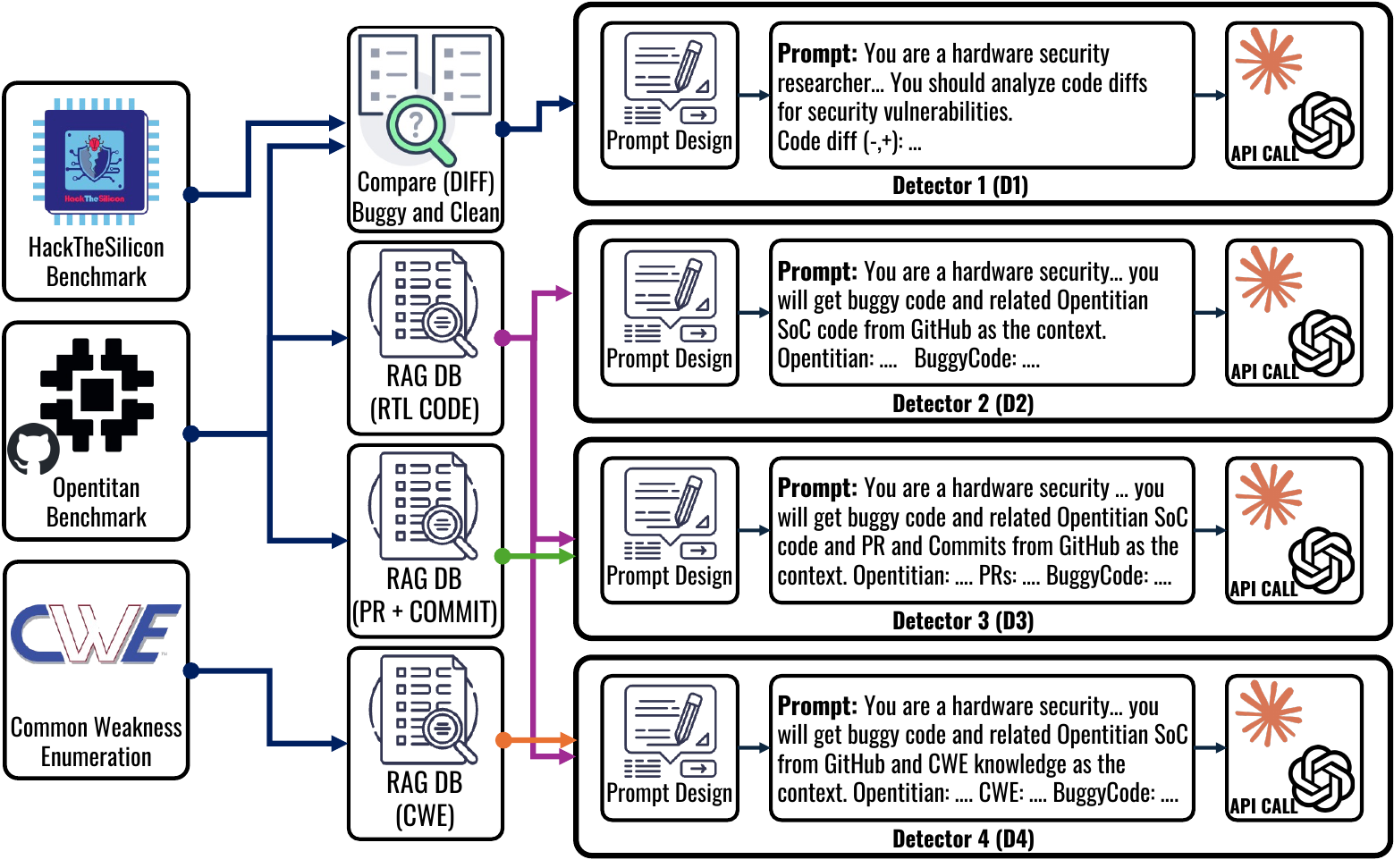}
    \caption{Overview of \ourname detectors.}
    \Description{A diagram showing different experiments conducted in \ourname.}
    \label{fig:detectors}
\end{figure}

Figure~\ref{fig:detectors} provides an overview of the experimental setup for different detectors. For each candidate program instance, regardless of detector type, the output requirements are standardized. Each detector must: (1) determine the presence of a vulnerability in the given source, (2) identify the lines responsible for the buggy behavior, and (3) produce a structured report explaining the defect mechanism, triggering conditions, and expected impact on program behavior. A critical distinction exists in the input modalities: D1 receives only diffs between buggy and original code, while D2-D4 receive complete buggy source code and leverage RAG systems to retrieve relevant context (e.g., RTL code and commits), representing typical LLM deployment patterns for code analysis.

\noindent\textbf{Diff-Aware LLM (D1).} This detector represents the straightforward detection approach. The detector receives only a unified diff output between the candidate buggy code and the reference baseline from the repository's latest commit. The LLM is directly prompted to analyze whether the given diff, i.e., \texttt{Code diff(-,+)}, introduces vulnerabilities and produces a structured report describing the defect security mechanism, triggering conditions, and expected impact. This detector directly tests whether simple syntactic diff analysis suffices for vulnerability detection without deeper code comprehension or security reasoning. 

\noindent\textbf{Repository-Aware RAG (D2).} The detector receives the complete buggy source code (not a diff), representing a realistic deployment where users of the LLM detection method have access to the entire vulnerable program. Before inference, the repository is indexed at the function-level granularity using code-aware chunking, which preserves boundaries and symbol information. At query time, a retrieval query is constructed from the buggy source code, and the RAG system retrieves relevant code fragments, function definitions, and module documentation from the indexed repository. This is the typical deployment pattern for LLM-powered code analysis tools ~\cite{guo2025svagent, collini2025marvel, saha2025sv, tarek2025bugwhisperer}. The model analyzes the buggy code in light of the retrieved context to assess the presence of vulnerabilities, identify the responsible lines, and ground its explanation in a broader understanding of the codebase. This detector assesses whether LLMs can identify vulnerabilities through genuine security analysis with repository context, without requiring diff hints.

\noindent\textbf{History-Aware RAG (D3).} Building on D2, this detector incorporates development history to test temporal reasoning. The RAG corpus is extended to include not only the SoC design but also relevant pull requests (PRs) and commit metadata, enabling the retrieval of discussions that motivated past changes to related components. Given the buggy code as input, the RAG system constructs queries spanning both code and history, retrieving relevant PR/commit alongside code fragments. The model was again prompted to assess the existence of any vulnerability in the given buggy code, considering the context. This detector assesses whether the historical development context enhances vulnerability reasoning beyond static code analysis.

\noindent\textbf{Weakness-Aware RAG (D4).} This detector represents the most sophisticated threat scenario with domain-specific security knowledge. The RAG system augments D2's repository context with a hardware security domain knowledge derived from the CWE database. At retrieval time, the system pairs repository context with concise CWE entries and implementation-oriented examples from the security domain, enabling the model to frame its diagnosis in terms of specific weakness classes. The model analyzes the buggy code against retrieved CWE patterns, determines the presence of bugs, highlights the responsible lines, and produces a report that labels the issue with CWE identifiers where applicable. This detector assesses whether domain-specific security knowledge facilitates genuine vulnerability reasoning, or whether LLMs continue to rely on pattern matching despite rich contextual support.

\begin{figure}
    \centering
    \includegraphics[width=\linewidth]{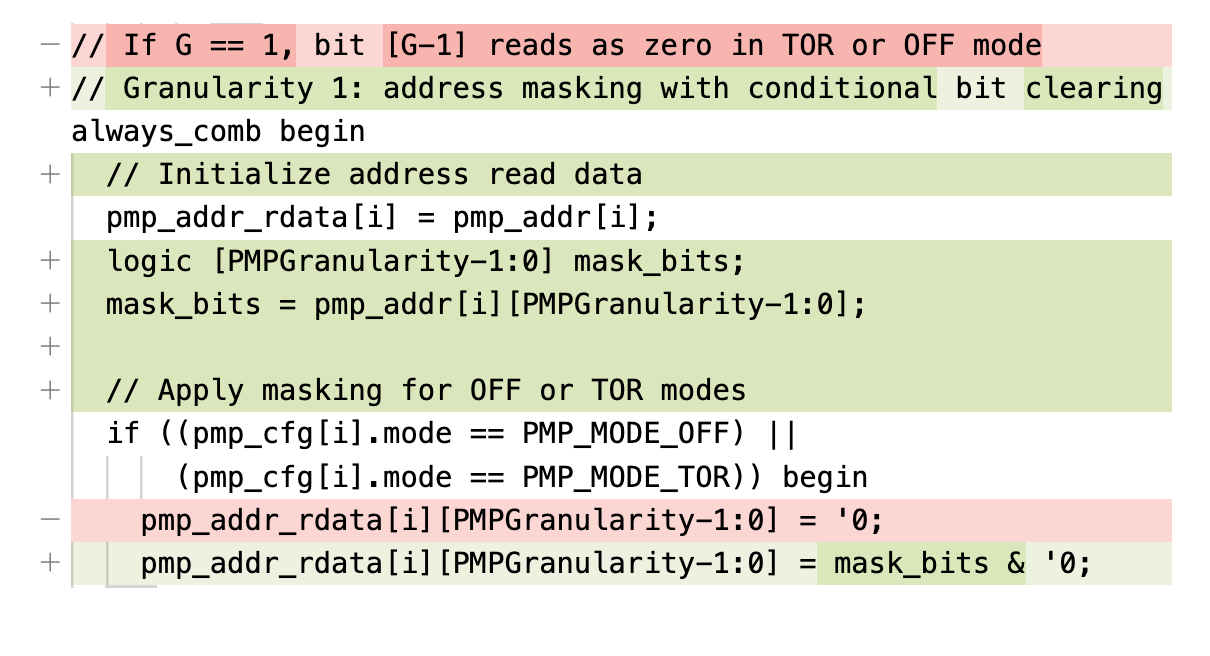}
    \caption{Example output from our obfuscation framework applied to the \texttt{ibex\_cs\_registers} module in OpenTitan~\cite{opentitan}. Additions are marked with green (+) and original replaced code is marked with red (–).}
    \Description{An Example of output of our obfuscation framwork on ibex_cs_registers modules of OpenTitan~\cite{openpiton}.}
    \label{fig:codediff}
\end{figure}

\begin{figure*}
    \centering
    \includegraphics[width=0.7\linewidth]{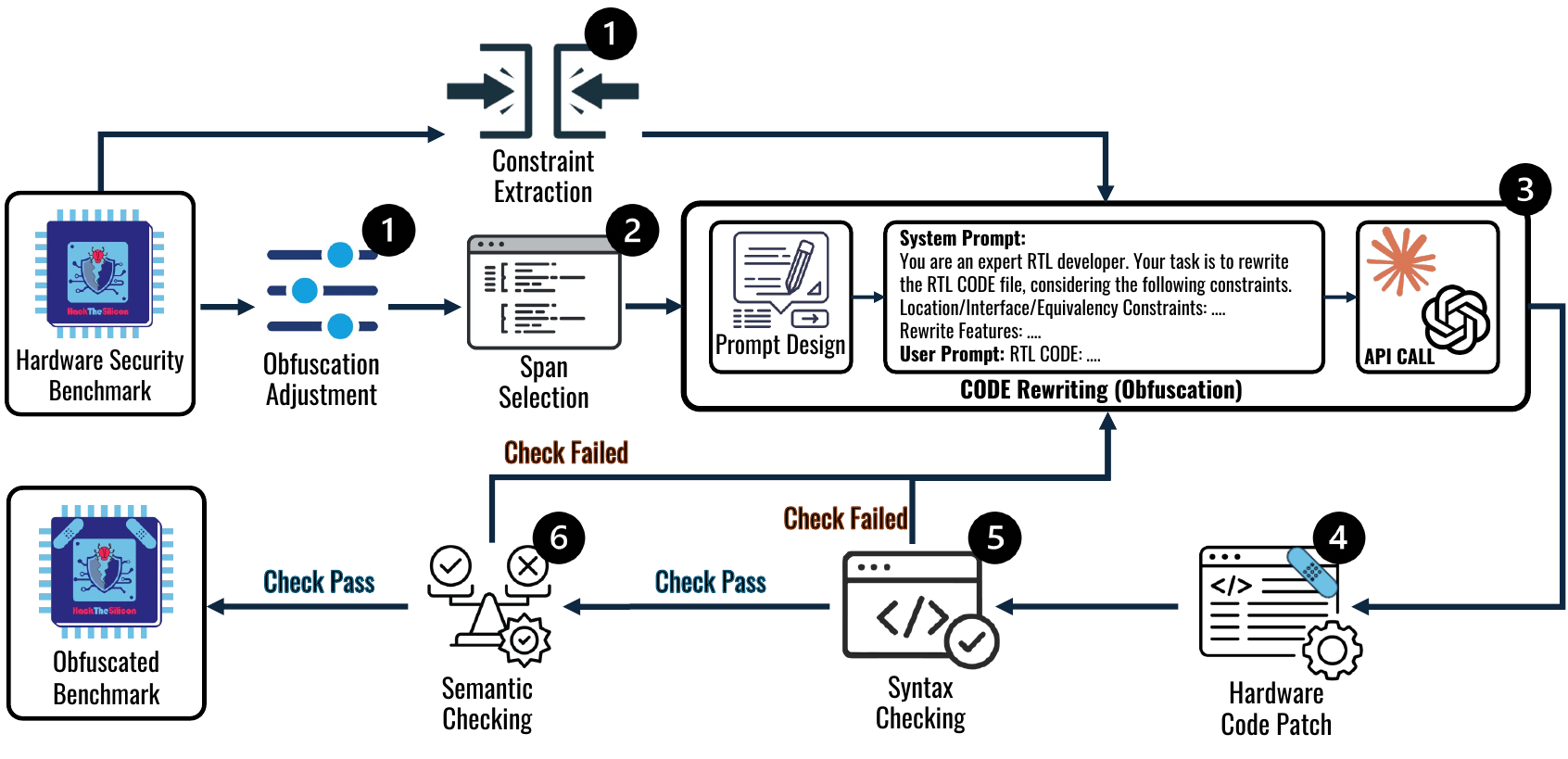}
    \caption{Overview of the proposed obfuscation framework.}
    \Description{A diagram showing the main components and workflow of the obfuscation framework.}
    \label{fig:overview}
\end{figure*}

\subsection{Obfuscation Framework}\label{subsec:method:obfuscation}

To enable a fair, research-grade benchmark that supports AI training, competition, and education while remaining suitable for evaluating LLM-based vulnerability detection methods, we introduce the first LLM-driven obfuscation framework for hardware security benchmarks. Given an existing industry-standard benchmark (e.g., \hackthesilicon~\cite{hackthesilicon}), our framework applies semantics-preserving transformations to security-critical hardware designs. Unlike conventional hardware obfuscation techniques intended for IP protection~\cite{aldecHdlObfuscation, semanticdesigns_verilog_obfuscator, verible}, our design explicitly targets LLM-based vulnerability analysis methods by preventing direct syntactic comparison (e.g., diff-based detection), thereby enabling the assessment of their true reasoning-based security analysis capabilities.

Figure~\ref{fig:overview} illustrates our introduced obfuscation pipeline.
The obfuscation process begins with \textbf{Constraint Extraction}~\circled{1}, where the framework instantiates formal invariants that the LLM must preserve during code rewriting. These constraints (1) prohibit modifications outside specified line ranges, (2) changes to module interfaces or port declarations, (3) alterations to signal widths or data types, (4) modifications to reset behavior, and (5) changes to finite-state machine semantics. These invariants are enforced through the rewriter's system prompt. The framework supports tunable obfuscation levels between 0\% to 100\% (\textbf{Obfuscation Adjustment}~\circled{1}), enabling controlled evaluation of robustness versus overhead (e.g., token usage). Based on the configured obfuscation level, the \textbf{Span Selection}~\circled{2} identifies vulnerable code regions to transform. The system verifies the validity of each span, ensuring no overlap with constraints and protected areas, and computes target boundaries with variations in line count.
This context enables the LLM to reason locally about equivalent rewrites while confining edits to the target range.

Next, for \textbf{Code Rewriting}~\circled{3}, the framework invokes a state-of-the-art LLM to perform functionally equivalent transformations using five strategies: (1) identifier renaming with semantically neutral alternatives, (2) control-flow restructuring through equivalent expressions and state machine reorganization, (3) diff-noise injection which disrupts syntactic similarity with canonical reference designs, (4) context dilution which obscures or disperses security-relevant cues and (5) logic rewriting through Boolean algebra and temporal transformations. Figure~\ref{fig:codediff} illustrates our obfuscation mechanisms through a representative OpenTitan code transformation. The LLM returns the rewritten segment as a \textbf{Hardware Code Patch}~\circled{4} with metadata (technique applied, confidence score, and potential side effects). A triviality filter strips comments and whitespaces to ensure that only meaningful edits are performed. \textbf{Syntax checking}~\circled{5} compiles the modified design using Verilator~\cite{verilator} for type checking, and SystemVerilog rule verification. If compilation fails, the framework captures error messages and provides them to the LLM for a retry attempt. This iterative refinement ensures all obfuscated code remains compilable. Beyond syntactic correctness, it is also crucial to ensure that the obfuscated code is functionally equivalent to the original source code. The \textbf{Semantic Checking}~\circled{6} stage performs this verification using the Yosys~\cite{yosys} tool, which uses SAT-based proofs for equivalence checking. If equivalence fails, the framework analyzes counterexample traces and restarts RTL rewriting with additional constraints derived from the failed proof to maintain benchmark integrity.

\section{Evaluation}\label{sec:evaluation}

Our evaluation investigates four fundamental questions. First, we assess \emph{how effective are current LLM-based vulnerability detectors in identifying hardware vulnerabilities when diff-based or reference-based shortcuts remain available?} (\textbf{RQ1}). Second,\emph{ we examine whether LLM-based vulnerability detection methods perform principled security reasoning about vulnerabilities?}
(\textbf{RQ2}). Third, \emph{we measure how effectively our obfuscation framework preserves human readability while preventing LLM-based vulnerability detection from reporting vulnerabilities based on diff analysis rather than genuine reasoning about vulnerability and security feature malfunctions} (\textbf{RQ3}). Finally, \emph{we analyze how different levels of obfuscation impact the accuracy of LLM-based vulnerability detection} (\textbf{RQ4}).

In the rest of the section, first we describe our evaluation setup in Section~\ref{sec:evasetup} and then answer each of our research questions based on our experiments, presented in Section~\ref{sec:rq1},~\ref{sec:rq23}, and~\ref{sec:rq4}.

\subsection{Evaluation Setup}
\label{sec:evasetup}

\noindent\textbf{Benchmark Characteristics.} We evaluate our framework on three recent \hackthesilicon competition benchmarks: Hack@DATE'25, Hack@DAC'25, and Hack@CHES'25.
All benchmarks are based on Google's OpenTitan~\cite{opentitan}, an open-source silicon root-of-trust platform incorporating cryptographic accelerators, secure boot, and key management. The benchmarks contain $50$ unique vulnerabilities across $27$ CWE categories, including top Hardware CWEs~\cite{cwehwtopten}. Vulnerabilities span multiple security domains including sensitive information in resource not removed before reuse (CWE-226~\cite{cwe226}), improper isolation of shared resources on SoC (CWE-1189~\cite{cwe1189}), 
on-chip debug and test interface with improper access control (CWE-1191~\cite{cwe1191}), 
improper handling of overlap between protected memory ranges (CWE-1260~\cite{cwe1260}), and improper access control for register interface (CWE-1262~\cite{cwe1262}). The vulnerabilities are distributed across $26$ hardware modules, including AES/HMAC engines, flash/OTP controllers, entropy sources, and the Ibex RISC-V core~\cite{ibex}.

\noindent\textbf{LLM Models.} We evaluate Claude Sonnet 4.5~\cite{claude45} and GPT-5~\cite{gpt5}, representing state-of-the-art LLMs for code analysis and reasoning~\cite{claude45,gpt5}. Both models accessed via API with $128,000$ token limits and temperature $0.0$ to maximum reproducibility.

\noindent\textbf{Verification Tools.} We employ Verilator~\cite{verilator} for syntax validation and Yosys~\cite{yosys} for formal semantic equivalence checking. Yosys performs RTL-to-RTL equivalence verification using \texttt{equiv\_make} and \texttt{equiv\_simple} commands with SAT-based proof generation.

\noindent\textbf{LLM-based Hardware Vulnerability Detector.} We implemented all four vulnerability detection methods described in Section~\ref{subsec:method:bugdetection} and evaluated their performance against our obfuscation framework (Section~\ref{subsec:method:obfuscation}) using four obfuscation levels (0\%, 10\%, 50\%, 100\%). Each experiment configuration is evaluated with three independent runs to account for LLM nondeterminism, using standardized prompts with temperature 0.0 and a 5,000-token output limit. For RQ2–RQ4, we focus on D1 and D4 as representative extremes: D1 isolates explicit syntactic comparison, while D4 represents the strongest RAG-enhanced detectors. Since preliminary runs showed D2/D3 exhibit similar trends to D4, evaluating D1 and D4 captures meaningful detector variation while managing experimental cost.

\noindent\textbf{Evaluation Metric.} We measure two primary performance metrics: vulnerability detection rate and accuracy. \textit{Detection Rate} quantifies the percentage of ground-truth injected vulnerabilities correctly identified by the detector, reflecting its recall capability. \textit{Detection Accuracy} measures the percentage of detector-reported vulnerabilities that correspond to actual vulnerabilities, capturing its precision and resistance to false positives. Together, these metrics characterize both the ability of a detector to fully find real vulnerabilities and its reliability in avoiding erroneous reports.

\subsection{RQ1: Baseline LLM Vulnerability Detection Effectiveness}
\label{sec:rq1}
\begin{figure}
    \centering
    \includegraphics[width=\linewidth]{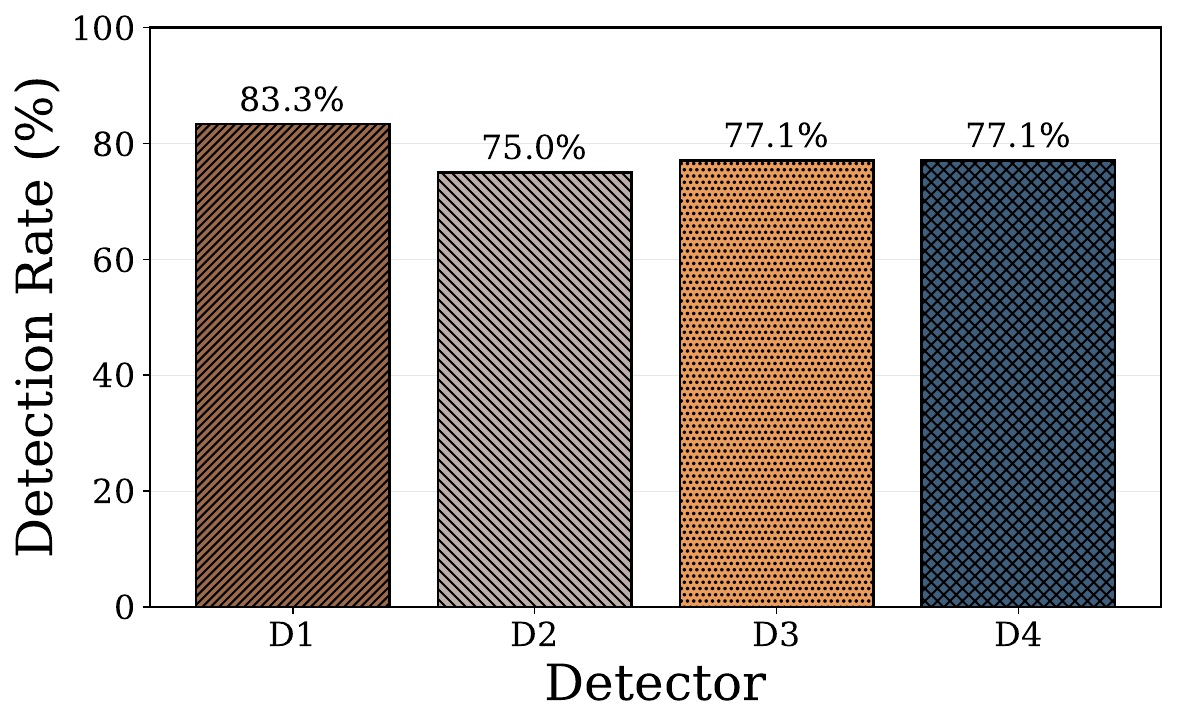}
    \caption{Effectiveness of LLM-based Vulnerability Detectors on \hackthesilicon~\cite{hackthesilicon} Benchmarks.}
    \Description{Effectiveness of Different  LLM-based Vulnerability Detectors on \hackthesilicon Benchmark.}
    \label{fig:rq1}
\end{figure}

We evaluate four detectors (D1–D4) on unobfuscated \hackthesilicon benchmarks~\cite{hackthesilicon}. 
Across both LLM models, we observe high vulnerability detection rates: D1 = 83.3\%, D2 = 75.0\%, D3 = 77.1\%, and D4 = 77.1\% (Figure~\ref{fig:rq1}). These results demonstrate that when syntactic or semantic reference information is accessible, either explicitly as in D1, or implicitly through contextual retrieval as in D2–D4, LLM-based detectors can identify approximately three out of four injected vulnerabilities, answering RQ1. 

However, the closely aligned performance between the direct diff-based approach (D1) and the context-retrieval approaches (D2–D4) suggests that all detectors employ similar internal strategies to identify vulnerabilities and raise the critical question: \textit{are LLMs performing genuine security analysis, or simply exploiting the availability of syntactically similar original code?} In real-world industry scenarios, verification tools have access to only a single codebase, there is no "original clean version" to compare against. If these LLM-based detectors are performing genuine security reasoning, they should maintain similar detection rates and accuracy even when the original code is not accessible. The following research questions (RQ2–RQ4) investigate this hypothesis by evaluating detector performance when our obfuscation framework simulates real-world conditions.

\begin{figure}
    \centering
    \includegraphics[width=\linewidth]{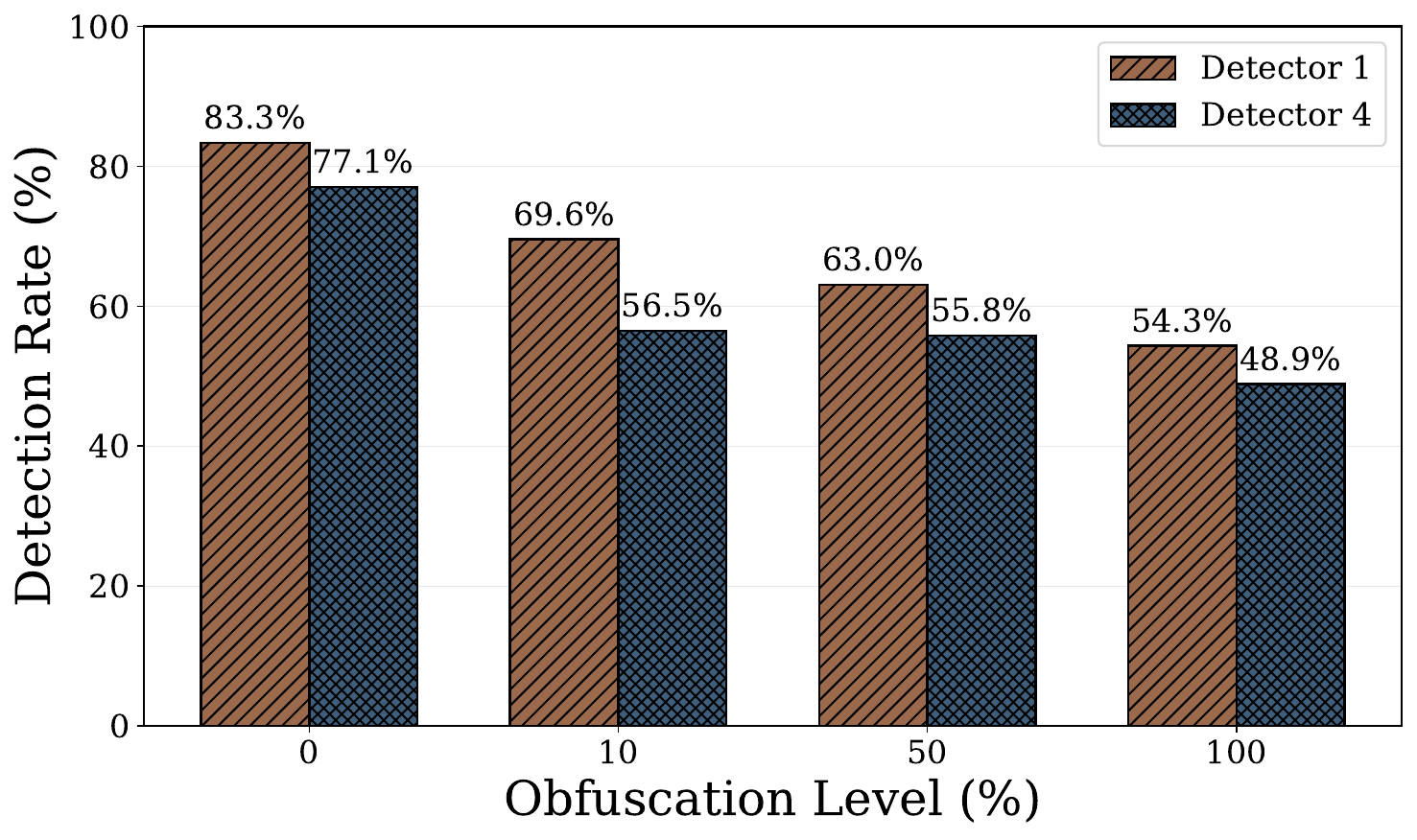}
    \caption{Effectiveness of Different Levels of Our Obfuscation 
    on LLM-based Vulnerability Detection Rate.}
    \Description{Effectiveness of Different Level of Obfuscation using \ourname on LLM-based Vulnerability Detection Rate.}
    \label{fig:detection_rate}
\end{figure}

\begin{figure}
    \centering
    \includegraphics[width=\linewidth]{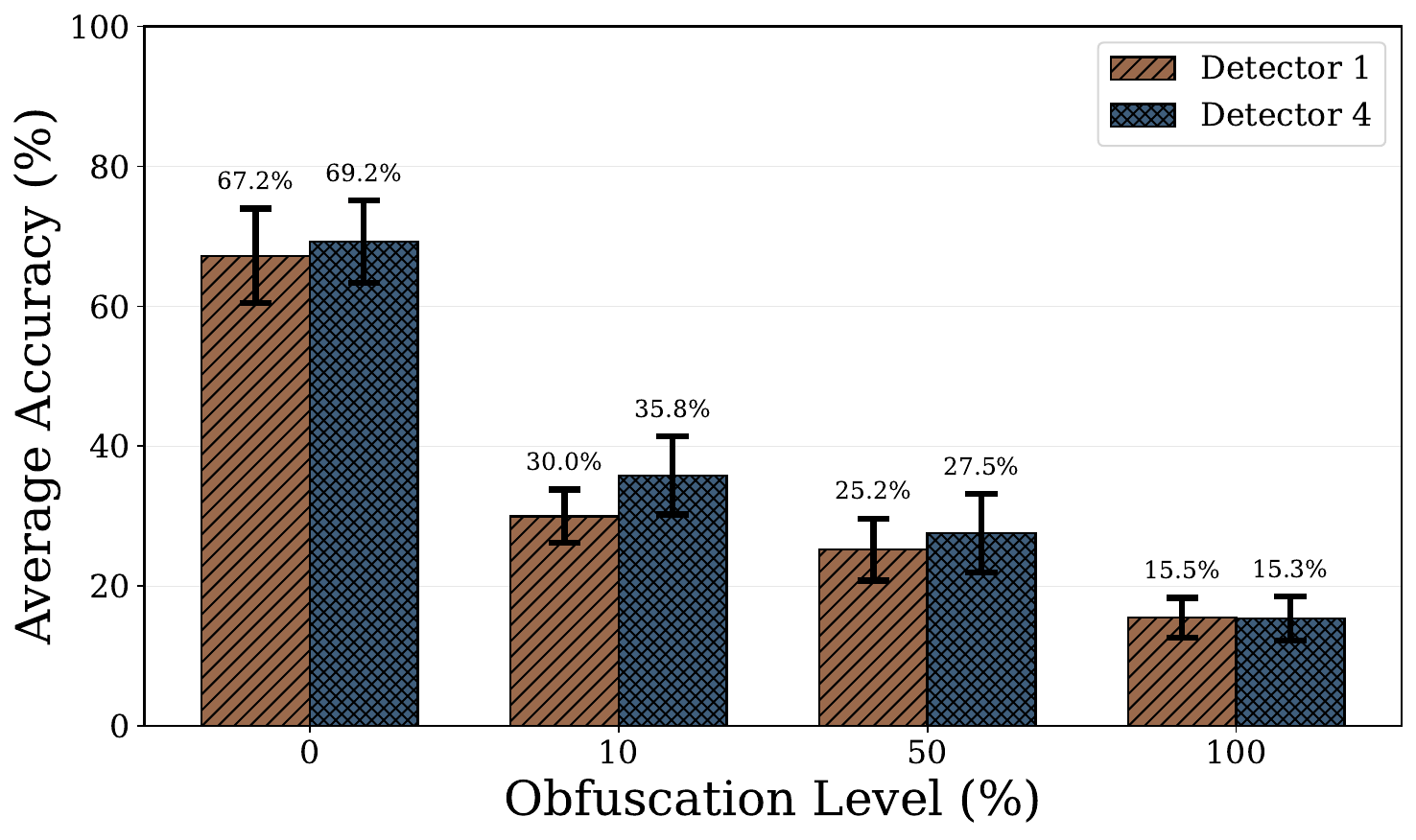}
    \caption{Effectiveness of Different Levels of Our Obfuscation 
    on LLM-based Vulnerability Detection Accuracy.}
    \Description{Effectiveness of Different Level of Obfuscation using \ourname on LLM-based Vulnerability Detection Accuracy.}
    \label{fig:accuracy}
\end{figure}

\subsection{RQ2 \& RQ3: Obfuscation Against Diff-Based Detection}
\label{sec:rq23}

Figure~\ref{fig:detection_rate} and ~\ref{fig:accuracy} present both detection rates and accuracy metrics across obfuscation levels for D1 and D4. While D1's detection rate decreases from 83.3\% to 54.3\% (35\% reduction) with 100\% obfuscation, its accuracy falls far more severely from 67.2\% to 15.5\% (77\% reduction). Similarly, D4 shows a decline in detection rate from 77.1\% to 48.9\% (37\% reduction), while accuracy falls from 69.2\% to 15.3\% (78\% reduction).

The substantial gap between detection and accuracy degradation reveals a critical insight: as syntactic changes increase, LLMs report more vulnerabilities, resulting in more false positives (FPs) and consequently decreasing accuracy. If LLMs performed genuine security analysis, semantic-preserving obfuscation should have minimal impact on the quality of detection. Instead, accuracy reductions far exceed detection rate reductions, empirically proving that LLM-based detectors focus on syntactic comparisons (diff-based detection) rather than actual vulnerability analysis. Notably, even D4, which receives RAG context rather than explicit diffs, exhibits this behavior, indicating that it ultimately performs diff-based analysis internally. This explains why increased syntactic differences between original and obfuscated benchmarks directly correlate with increased false positives, a pattern inconsistent with genuine security reasoning but consistent with diff-based exploitation.

\subsection{RQ4: Impact of Obfuscation Levels on LLM Detection}
\label{sec:rq4}
Analyzing progressive degradation across obfuscation levels, Figure~\ref{fig:detection_rate} and ~\ref{fig:accuracy}, reveals that detection rates decrease monotonically (D1: 83.3\% $\rightarrow$ 54.3\%; D4: 77.1\% $\rightarrow$ 48.9\%), while accuracy shows dramatic non-linear degradation with 48-55\% drops at just 10\% obfuscation. The RAG-enhanced detector (D4) demonstrates greater initial sensitivity (21\% detection drop vs. 16\% for D1 at 10\%), suggesting RAG retrieval mechanisms are particularly vulnerable to syntactic transformation.
Our results illustrated, even minimal 10\% obfuscation using our method provides substantial accuracy reduction (48-55\%), while 100\% achieves near-complete mitigation (77-78\% reduction).

\section{Related Work}\label{sec:related}

Software obfuscation has extensive literature~\cite{barak2016hopes, collberg1997taxonomy, xu2020layered}. Techniques include control flow obfuscation~\cite{balachandran2014function}, data obfuscation~\cite{bakken2004data}, and opaque predicates~\cite{xu2016generalized}. Hardware obfuscation primarily focuses on IP protection.

\noindent\textbf{Traditional Hardware IP Protection Tools.} The hardware security community has developed numerous obfuscation approaches to protect intellectual property from unauthorized disclosure and reverse engineering. Commercial solutions such as Aldec's HDL Code Obfuscator~\cite{aldecHdlObfuscation} and Semantic Designs' Verilog Obfuscator~\cite{semanticdesigns_verilog_obfuscator} employ systematic identifier renaming using cryptographic hashes, structural information removal, and formatting elimination to maximize protection against IP theft. The Verible toolchain~\cite{verible} provides similar capabilities through automated symbol mangling and semantic cue elimination. While these tools effectively protect proprietary designs, their aggressive obfuscation strategies—including complete identifier encryption and structural anonymization, render code unreadable to human and AI analysts, fundamentally incompatible with educational and AI learning objectives that require comprehensible security analysis.

\noindent \textbf{LLM-Based Gate-Level Obfuscation.} Recent research has explored leveraging LLMs for automated hardware logic obfuscation~\cite{latibari2024automated, gandhi2024large}. These approaches operate at the gate-level netlist abstraction, instructing LLMs to insert key-programmable logic gates (XOR/XNOR functions), multiplexers, and dummy circuit elements while preserving input-output interface specifications. The resulting obfuscated netlists maintain functional correctness but require sophisticated reverse engineering to extract the original design intent. However, gate-level obfuscation fundamentally transforms the analysis domain from RTL code examination to netlist reconstruction. This eliminates the core human and AI learning objective of hardware security competitions and benchmarks.

Despite their effectiveness for IP protection, existing hardware obfuscation approaches are fundamentally unsuitable for creating a hardware security benchmark that can be used for training machine learning models, participating in capture-the-flag competitions, and evaluating LLM and AI-based vulnerability detection approaches. Traditional IP protection tools prioritize complete code obscuration over human and AI comprehensibility, employing encryption, aggressive symbol mangling, and structural elimination that prevent meaningful security analysis by both humans and AI. These limitations motivate our development of the first LLM-assisted obfuscation, designed explicitly to create benchmarks with unique considerations for human and AI learning settings, and to evaluate AI-based hardware vulnerability detection methods while preventing automated diff-based exploitation of LLM-based approaches.

\section{Conclusion}\label{sec:conclusion}
This paper presented the first systematic analysis revealing that LLM-based hardware vulnerability detection approaches primarily rely on diff-based pattern matching rather than genuine security reasoning when evaluated on open-source benchmarks like \hackthesilicon, achieving 75-83\% detection rates through automated exploitation of diff operations against known original designs. In response, we introduced the first LLM-based obfuscation framework specifically designed for hardware security competitions and benchmarks, employing semantic-preserving transformations that prevent diff-based exploitation while maintaining human readability. Experimental results demonstrate substantial effectiveness: even minimal 10\% obfuscation reduces LLM accuracy by 48-55\%, while comprehensive 100\% obfuscation achieves maximum detection rate and accuracy reductions of 55\% and 78\% respectively, reducing final accuracy to approximately 15\%, successfully disadvantaging diff-based approaches while maintaining functionality.

\begin{acks}
Our research was partially funded by Intel’s Scalable Assurance Program, DFG-SFB 1119-236615297, NSF-DFG-Grant 538883423, and the ERC Programme-Grant 101055025-HYDRANOS. The views expressed in this work are those of the authors and do not represent endorsements or official positions of the funding organizations.
\end{acks}

\clearpage

\printbibliography

@article{ahmad2025lashed,
  title={LASHED: LLMs And Static Hardware Analysis for Early Detection of RTL Bugs},
  author={Ahmad, Baleegh and Pearce, Hammond and Karri, Ramesh and Tan, Benjamin},
  journal={arXiv preprint arXiv:2504.21770},
  year={2025}
}

@proceedings{10.1145/3316781,
title = {DAC '19: Proceedings of the 56th Annual Design Automation Conference 2019},
year = {2019},
isbn = {9781450367257},
publisher = {Association for Computing Machinery},
address = {New York, NY, USA},
location = {Las Vegas, NV, USA}
}

@article{akter2023survey,
  title={A Survey on Hardware Security: Current Trends and Challenges},
  author={Akter, Sonia and Khalil, Kasem and Bayoumi, Magdy},
  journal={IEEE Access},
  volume={11},
  pages={77543--77565},
  year={2023},
}

@misc{aldecHdlObfuscation,
  author       = {{Aldec, Inc.}},
  title        = {{HDL Code Obfuscation}},
  howpublished = {\url{https://www.aldec.com/en/support/resources/documentation/articles/1586}},
  note         = {Accessed: 2025-11-10}
}

@misc{claude45,
 	title     =  {{Introducing Claude Sonnet 4.5}},
 	author    =  {{Anthropic}},
    year      = {2025},
 	howpublished =  {\url{https://www.anthropic.com/news/claude-sonnet-4-5}},
     note = {Online; accessed 2025-11-10}
}

@article{bakken2004data,
  title={Data obfuscation: Anonymity and desensitization of usable data sets},
  author={Bakken, David E and Rarameswaran, R and Blough, Douglas M and Franz, Andy A and Palmer, Ty J},
  journal={IEEE Security \& Privacy},
  volume={2},
  number={6},
  pages={34--41},
  year={2004},
  publisher={IEEE}
}

@inproceedings{balachandran2014function,
  title={Function level control flow obfuscation for software security},
  author={Balachandran, Vivek and Keong, Ng Wee and Emmanuel, Sabu},
  booktitle={2014 Eighth International Conference on Complex, Intelligent and Software Intensive Systems},
  pages={133--140},
  year={2014},
  organization={IEEE}
}

@article{barak2016hopes,
  title={Hopes, fears, and software obfuscation},
  author={Barak, Boaz},
  journal={Communications of the ACM},
  volume={59},
  number={3},
  pages={88--96},
  year={2016},
  publisher={ACM New York, NY, USA}
}

@inproceedings{chen2022trusting,
  title={Trusting the Trust Anchor: Towards Detecting Cross-Layer Vulnerabilities With Hardware Fuzzing},
  author={Chen, Chen and Kande, Rahul and Mahmoody, Pouya and Sadeghi, Ahmad-Reza and Rajendran, JV},
  booktitle={Proceedings of the 59th ACM/IEEE Design Automation Conference},
  pages={1379--1383},
  year={2022}
}

@misc{verible,
  title        = {Verible: A Suite of SystemVerilog Developer Tools},
  howpublished = {\url{https://github.com/chipsalliance/verible}},
  author       = {{Chipsalliance}},
  year         = {2025},
  note         = {Accessed: 2025-11-10}
}

@misc{collberg1997taxonomy,
  title={A taxonomy of obfuscating transformations},
  author={Collberg, Christian and Thomborson, Clark and Low, Douglas},
  year={1997},
  publisher={Technical Report 148, Department of Computer Science, University of Auckland}
}

@article{collini2025marvel,
  title={MARVEL: Multi-Agent RTL Vulnerability Extraction using Large Language Models},
  author={Collini, Luca and Ahmad, Baleegh and Ah-kiow, Joey and Karri, Ramesh},
  journal={arXiv preprint arXiv:2505.11963},
  year={2025}
}

@article{dessouky2019hardfails,
  title={{HardFails: Insights into Software-Exploitable Hardware Bugs}},
  author={Dessouky, Ghada and Gens, David and Haney, Patrick and Persyn, Garrett and Kanuparthi, Arun and Khattri, Hareesh and Fung, Jason M and Sadeghi, Ahmad-Reza and Rajendran, Jeyavijayan},
  journal={USENIX Security Symposium},
  year={2019}
}

@inproceedings{gandhi2024large,
  title={Large language model driven logic locking: A generative approach to secure ic design},
  author={Gandhi, Jugal and Shekhawat, Diksha and Santosh, M and Dofe, Jaya and Pandey, Jai Gopal},
  booktitle={2024 IEEE 33rd Asian Test Symposium (ATS)},
  pages={1--4},
  year={2024},
  organization={IEEE}
}

@article{guo2025svagent,
  title={SVAgent: AI Agent for Hardware Security Verification Assertion},
  author={Guo, Rui and Ayalasomayajula, Avinash and Li, Henian and Zhou, Jingbo and Saha, Sujan Kumar and Farahmandi, Farimah},
  journal={arXiv preprint arXiv:2507.16203},
  year={2025}
}

@misc{hackdac_origin,
  title={Raising Awareness of Hardware Security Weaknesses: Intel Research and Hack@DAC},
  author={{Intel}},
  year={2025},
  howpublished={\url{https://www.intel.com/content/www/us/en/security/security-practices/blogs/raising-awareness-hardware-security-weaknesses.html}},
  note         = {Accessed: 2025-11-05}
}

@misc{Kanuparthi2024HackAtDAC,
  author    = {Arun Kanuparthi and Hareesh Khattri and Jason Fung and J. V. Rajendran and Ahmad-Reza Sadeghi and Rahul Kande and Chen Chen and Mohamadreza Rostami},
  title     = {The Hack@DAC Story: Learnings from Organizing the World's Largest Hardware Hacking Competition},
  howpublished = {Presented at Black Hat USA},
  year      = {2024},
  note      = {Briefings Presentation},
  url       = {https://i.blackhat.com/BH-US-24/Presentations/US-24-Kanuparthi-HackAtDACStory-Wednesday.pdf}
}

@article{koylu2023survey,
  title={A survey on machine learning in hardware security},
  author={K{\"o}yl{\"u}, Troya {\c{C}}a{\u{g}}{\i}l and Wedig Reinbrecht, Cezar Rodolfo and Gebregiorgis, Anteneh and Hamdioui, Said and Taouil, Mottaqiallah},
  journal={ACM Journal on Emerging Technologies in Computing Systems},
  volume={19},
  number={2},
  pages={1--37},
  year={2023},
  publisher={ACM New York, NY}
}

@inproceedings{latibari2024automated,
  title={Automated hardware logic obfuscation framework using gpt},
  author={Latibari, Banafsheh Saber and Ghimire, Sujan and Chowdhury, Muhtasim Alam and Nazari, Najmeh and Gubbi, Kevin Immanuel and Homayoun, Houman and Sasan, Avesta and Salehi, Soheil},
  booktitle={2024 IEEE 17th Dallas Circuits and Systems Conference (DCAS)},
  pages={1--5},
  year={2024},
  organization={IEEE}
}

@misc{ibex,
 	title     =  {{lowRISC ibex RISC-V Core}},
 	author    =  {{lowRISC}},
 	howpublished =  {\url{https://github.com/lowRISC/ibex}},
    note = {Online; accessed 2025-11-10}
}

@article{marks2025ResearchGaps,
  author    = {Marks, Paul},
  title     = {Security Research Gaps Leave Critical Infrastructure Open to Cyberattack},
  journal   = {Communications of the ACM (CACM) News},
  year      = {2025},
  url       = {https://cacm.acm.org/news/security-research-gaps-leave-critical-infrastructure-open-to-cyberattack/}
}

@article{matarazzo2025survey,
  title={A Survey on Large Language Models With Some Insights on Their Capabilities and Limitations},
  author={Matarazzo, Andrea and Torlone, Riccardo},
  journal={arXiv preprint arXiv:2501.04040},
  year={2025}
}

@misc{mitre,
	title     =  {Common Weakness Enumeration},
	author    =  {MITRE},
	url =  {https://cwe.mitre.org},
    note = {Online; accessed 2025-11-10}
}

@misc{cwehwtopten,
 	title     =  {2025 Most Important Hardware Weaknesses},
 	author    =  {MITRE},
    year = {2025},
    url = {https://cwe.mitre.org/topHW/archive/2025/2025_CWE_MIHW.html},
    note = {Online; accessed 2025-11-10}
}

@misc{cwe1189,
  author  = {{MITRE}},
  title   = {CWE-1189: Improper Isolation of Shared Resources on System-on-a-Chip (SoC)},
  url     = {https://cwe.mitre.org/data/definitions/1189.html},
  note = {Online; accessed 2025-11-10}
}

@misc{cwe1191,
    author    =  {{MITRE}},
 	title     =  "CWE-1191: On-Chip Debug and Test Interface With Improper Access Control",
 	url =  {https://cwe.mitre.org/data/definitions/1191.html},
    note = {Online; accessed 2025-11-10}
}

@misc{cwe1260,
 	title     =  {{CWE-1260: Improper Handling of Overlap Between Protected Memory Ranges}},
 	author    =  {{MITRE}},
 	url =  {https://cwe.mitre.org/data/definitions/1260.html},
    note = {Online; accessed 2025-11-10}
}

@misc{cwe1262,
 	title     =  {{CWE-1262: Improper Access Control for Register Interface}},
 	author    =  {{MITRE}},
 	url =  {https://cwe.mitre.org/data/definitions/1262.html},
    note = {Online; accessed 2025-11-10}
}

@misc{cwe226,
    title = {{CWE-226: Sensitive Information in Resource Not Removed Before Reuse}},
    author = {{MITRE}},
    url = {https://cwe.mitre.org/data/definitions/226.html},
    note = {Online; accessed 2025-11-10}
}

@misc{opentitan,
  author       = {{OpenTitan Project}},
  title        = {OpenTitan: Open Source Silicon Root of Trust},
  url          = {https://opentitan.org},
  note         = {Accessed: 2025-11-12}
}

@misc{gpt5,
 	title     =  {{GPT-5 is here}},
 	author    =  {{OpenAI}},
 	url       =  {https://openai.com/gpt-5/},
    note = {Online; accessed 2025-11-10}
}

@misc{semanticdesigns_verilog_obfuscator,
  author       = {{Semantic Designs}},
  title        = {{Verilog Source Code Obfuscator}},
  url = {https://semanticdesigns.com/Products/Obfuscators/VerilogObfuscator.html},
  note         = {Accessed: 2025-11-10}
}

@article{balkind2016openpiton,
  title={OpenPiton: An open source manycore research framework},
  author={Balkind, Jonathan and McKeown, Michael and Fu, Yaosheng and Nguyen, Tri and Zhou, Yanqi and Lavrov, Alexey and Shahrad, Mohammad and Fuchs, Adi and Payne, Samuel and Liang, Xiaohua and others},
  journal={ACM SIGPLAN Notices},
  volume={51},
  number={4},
  pages={217--232},
  year={2016},
  publisher={ACM New York, NY, USA}
}

@article{rostami2024fuzzerfly,
  title={{Fuzzerfly Effect: Hardware Fuzzing for Memory Safety}},
  author={Rostami, Mohamadreza and Chen, Chen and Kande, Rahul and Li, Huimin and Rajendran, Jeyavijayan and Sadeghi, Ahmad-Reza},
  journal={IEEE Security \& Privacy},
  year={2024}
}

@article{specure,
    author = {Rostami, Mohamadreza and Zeitouni, Shaza and Kande, Rahul and Chen, Chen and Mahmoody, Pouya and Rajendran, Jeyavijayan and Sadeghi, Ahmad-Reza},
    title = {{Lost and Found in Speculation: Hybrid Speculative Vulnerability Detection}},
    journal={ACM/IEEE Design Automation Conference},
    year = {2024}
}

@inproceedings{sadeghi2021organizing,
  title={Organizing The World's Largest Hardware Security Competition: Challenges, Opportunities, and Lessons Learned},
  author={Sadeghi, Ahmad-Reza and Rajendran, Jeyavijayan and Kande, Rahul},
  booktitle={Proceedings of the 2021 Great Lakes Symposium on VLSI},
  pages={95--100},
  year={2021}
}

@article{saha2025sv,
  title={SV-LLM: An Agentic Approach for SoC Security Verification using Large Language Models},
  author={Saha, Dipayan and Tarek, Shams and Shaikh, Hasan Al and Hasan, Khan Thamid and Nalluri, Pavan Sai and Hasan, Md Ajoad and Alam, Nashmin and Zhou, Jingbo and Saha, Sujan Kumar and Tehranipoor, Mark and others},
  journal={arXiv preprint arXiv:2506.20415},
  year={2025}
}

@misc{synopsys,
 	title     =  {{VCS: Functional Verification Solution}},
 	author    =  {{Synopsys}},
 	url =  {https://www.synopsys.com/verification/simulation/vcs.html},
     note = {Online; accessed 2025-11-10}
}

@misc{verilator,
 	title     =  {{Verilator, the fastest Verilog/SystemVerilog simulator.}},
 	author    =  {{Veripool}},
 	url =  {https://www.veripool.org/verilator/},
     note = {Online; accessed 2025-11-10}
}

@inproceedings{tarek2025bugwhisperer,
  title={Bugwhisperer: Fine-tuning llms for soc hardware vulnerability detection},
  author={Tarek, Shams and Saha, Dipayan and Saha, Sujan Kumar and Farahmandi, Farimah},
  booktitle={2025 IEEE 43rd VLSI Test Symposium (VTS)},
  pages={1--5},
  year={2025},
  organization={IEEE}
}

@misc{hackthesilicon,
  title={Hack The Silicon},
  author={HackTheSilicon Team},
  url={https://hackthesilicon.com/},
  note = {Online; accessed 2025-11-10}
}

@inproceedings{xu2016generalized,
  title={Generalized dynamic opaque predicates: A new control flow obfuscation method},
  author={Xu, Dongpeng and Ming, Jiang and Wu, Dinghao},
  booktitle={International Conference on Information Security},
  pages={323--342},
  year={2016},
  organization={Springer}
}

@article{xu2020layered,
  title={Layered obfuscation: a taxonomy of software obfuscation techniques for layered security},
  author={Xu, Hui and Zhou, Yangfan and Ming, Jiang and Lyu, Michael},
  journal={Cybersecurity},
  volume={3},
  number={1},
  pages={9},
  year={2020},
  publisher={Springer}
}

@misc{yosys,
  title        = {Yosys: A Suite of Open-Source RTL Synthesis Tools},
  author       = {{YosysHQ}},
  url = {https://github.com/YosysHQ/yosys},
  year         = {2025},
  note = {Online; accessed 2025-11-10}
}

\end{document}